\font\caps=cmcsc10 at 12pt
\font\eightrm=cmr8 at 8pt
\documentclass{oxarticle}
\usepackage{amsmath, amssymb}
\usepackage{graphics, setspace}
\usepackage{epstopdf}

\newcommand{\exinv}{exotic invariant}

\newcommand{\cE}{{\cal E}}

\newcommand{\cB}{{\cal B}}
\newcommand{\cG}{{\cal G}}

\newcommand{\cA}{{\cal A}}

\newcommand{\cR}{{\cal R}}

\newcommand{\bt}{\begin{tabular}{c}}
\newcommand{\et}{\end{tabular}}

\newcommand{\eb}{\ee\be } 
\newcommand{\ebp}{\rt.\ee\be\lt.} 
\newcommand{\bmat}{\lt ( \begin{array} }
\newcommand{\emat}{  \end{array} \rt )}

\newcommand{\oH}{{\ov H}}
\newcommand{\oP}{{\ov P}}

\newcommand{\oQ}{{\ov Q}}

\newcommand{\cH}{{\cal H}}
\newcommand{\cP}{{\cal P}}

\newcommand{\ovD}{{\ov D}}

\newcommand{\oJ}{{\ov J}}

\newcommand{\oq}{{\ov \q}}
\newcommand{\oT}{{\ov T}}
\newcommand{\oG}{{\ov \G}}

\newcommand{\oK}{{\ov K}}

\newcommand{\oY}{{\ov Y}}
\newcommand{\og}{{\ov g}}
\newcommand{\oy}{{\ov \y}}

\newcommand{\om}{{\ov m}}

\newcommand{\oC}{{\ov C}}

\newcommand{\oF}{{\ov F}}
\newcommand{\A}{{\ov A}}

\renewcommand{\a}{\alpha}	
\renewcommand{\b}{\beta}
\newcommand{\g}{\gamma}
\renewcommand{\d}{\delta}
\newcommand{\e}{\epsilon}

\newcommand{\q}{\theta}

\newcommand{\m}{\mu}
	
\newcommand{\x}{\xi}

\newcommand{\s}{\sigma}

\renewcommand{\t}{\tau}

\newcommand{\f}{\phi}

\newcommand{\y}{\psi}
\newcommand{\w}{\omega}
\newcommand{\G}{\Gamma}
\newcommand{\D}{\Delta}
	
\newcommand{\Lam}{\Lambda}
\newcommand{\X}{\Xi}
\renewcommand{\P}{\Pi}	
\renewcommand{\S}{\Sigma}

\newcommand{\F}{\Phi}	
\newcommand{\Y}{\Psi} 

\newcommand{\W}{\Omega}
\newcommand{\la}{\label}
\newcommand{\ci}{\cite}

\newcommand{\ds}{\documentstyle}	
\newcommand{\fr}{\frac}

\newcommand{\na}{\nabla}
\newcommand{\pa}{\partial}
\newcommand{\ov}{\overline}
\newcommand{\br}{\begin{rant}}
\newcommand{\er}{\end{rant}}
\newcommand{\beC}{\begin{Conjecture}}
\newcommand{\eeC}{\end{Conjecture}}
\newcommand{\be}{\begin{equation}}
\newcommand{\ee}{\end{equation}}
\newcommand{\ba}{\begin{array}} 
\newcommand{\ea}{\end{array}}
\newcommand{\bea}{\begin{eqnarray}}
\newcommand{\eea}{\end{eqnarray}}
\newcommand{\ra}{\rightarrow}
\newcommand{\Ra}{\Rightarrow}
\newcommand{\lra}{\longrightarrow}

\newcommand{\lt}{\left}
\newcommand{\rt}{\right}

\newcommand{\ben}{\begin{enumerate}}
\newcommand{\een}{\end{enumerate}}
\newcommand{\bitem}{\begin{itemize}}
\newcommand{\eitem}{\end{itemize}}

\newcounter{orange} 
\newcounter{apple} 
\addtocounter{apple}{1}

\newcounter{grape} 
\addtocounter{grape}{1}
\newcommand{\numberhere}{2024}
\newcommand{\articlenumber}{64SusyExoLett}

\usepackage{color}

\newcommand{\mathsym}[1]{{}}
\newcommand{\unicode}[1]{{}}

\begin{document}

 \begin{center}

{ \LARGE    Supersymmetry anomalies, exotic pairs and the  supersymmetric standard model }  
\vspace*{.1in}
%

\renewcommand{\thefootnote}{\fnsymbol{footnote}}

{\caps John A. Dixon\footnote{cybersusy @gmail.com}\\Physics Dept\\University of Calgary 
} \\[.5cm] 
\end{center}

\normalsize
\large
 \begin{center}
 Abstract 
\end{center}

For 34 years it has been known that chiral supersymmetry (SUSY) in 3+1 dimensions has a very large Becchi-Rouet-Stora (BRS)  cohomology space at ghost charge one. This suggests that there might be corresponding SUSY anomalies coming from linearly divergent, triangle, Feynman diagrams.  This paper discusses some progress and some outstanding issues related to these questions. The concept of exotic pairs is introduced, with an example from the massless supersymmetric standard model.


\refstepcounter{orange}
{\bf \theorange}.\;
{\bf  The well known anomalies:} 
\la{firstpara} 		Anomalies are well known to occur for various gauge theories in 3+1 dimensions \ci{weinberg2}. They play many roles, including the mediation of the decay $\P^0 \ra 2 \g$ \ci{pi2gamma}. The quantum numbers in the standard model appear to reflect a need to eliminate the gauge anomalies so as to preserve  unitarity and gauge invariance \ci{adler}. 

\refstepcounter{orange}
{\bf \theorange}.\;
{\bf   The anomalies of gauge theories are typically generated by  triangle Feynman diagrams that are linearly divergent.}  The anomalies are a  violation of  the gauge symmetry, and they cannot be removed by renormalization of the action. 
This impossibility of removing them is equivalent to the statement that anomalies are in the ghost charge one cohomology of the relevant nilpotent BRS operator \ci{brs,ZJ}. 
 The simplest example of an anomaly occurs in the conservation of axial charge for the well known gauge transformation 
\be
\d A_{\m} = \pa_{\m} \w.
\la{abeliangauge}
\ee
 Yang-Mills theory  \ci{brs} and   gravity (the latter in higher dimensions than 3+1)  \ci{witten} have similar anomalies.  These gauge   and gravity anomalies can also occur in supersymmetric theories, but these are not supersymmetry anomalies: they are gauge anomalies in a supersymmetric
 theory.

\refstepcounter{orange}
{\bf \theorange}.\;
{\bf   SUSY anomalies and exotic invariants:}  It has been generally accepted that supersymmetry is not anomalous \ci{noanom}.  However this paper disagrees with that view. As shown below, supersymmetry anomalies could occur for exotic invariants. These are well hidden, and discoverable only by using spectral sequences \ci{mccleary,dixspecseq}.  

\refstepcounter{orange}
{\bf \theorange}.\;
\la{puzzlepara}
{\bf  The puzzle of anomalies in SUSY:}  The first indications that there could be SUSY  anomalies arose from looking at the BRS cohomology of the SUSY transformations \ci{holes,holescommun} in the model of Wess and Zumino \ci{WZ}. But that created a puzzle: we could see the possibility of anomalies, but that raised the question of how to construct the operators that could be  anomalous.  
For SUSY, all the ghost charge one cohomology has uncontracted spinor indices, as in paragraph \ref{simplestanomaly} below.  No corresponding ghost charge zero BRS cohomology with uncontracted indices could be constructed by using the usual superspace methods.  So these questions have generated little interest, because the way to make them Lorentz invariant has  also 
been problematic.  The known gauge and gravity anomalies are all Lorentz invariant with contracted spinor and vector indices.   
Both of these questions are resolved in paragraph \ref{exoticpairpara} below.

\refstepcounter{orange}
{\bf \theorange}.\;{\bf  Ghost Charge:}  The ghost charge is found to start with here, by counting the number of ghosts $C_{\a}$ in an expression.  When we encounter the action in paragraph \ref{actionsec}, we need to recognize that $\G,Y,\Lam$ all have ghost charge -1.

\refstepcounter{orange}
{\bf \theorange}.\;
{\bf  `Exotic Pairs'. } 
\la{exoticpairpara}
  The answer to the problems in paragraph \ref{puzzlepara}  is to construct  `exotic invariants' that depend on antifields and which are also not superspace invariant, and to couple them to  chiral dotted spinor superfield sources.   
This results in Lorentz invariant `exotic pairs' of composite local operators of ghost charge zero and one, which might generate interesting SUSY anomalies.   A complicating factor is that there are also constraints, coming from the superpotential, that both members of each exotic pair must satisfy. The constraints are derived, using the spectral sequence, in paragraph \ref{tricky}. The simplest known solutions to these constraints can be found in the massless supersymmetric standard model (SSM).   They depend on the left/right assymetry in that model, as one can see from equation (\ref{electron}).

\refstepcounter{orange}
{\bf \theorange}.\;
\la{anomops}
{\bf  Examples of the  exotic SUSY invariants:} A reader who is familiar with the notation can look at paragraph \ref{simplestsourceofanom} for the simplest version of an exotic invariant.    The summary in paragraph \ref{summary} below provides  a simple explanation of the unusual features of these  exotic SUSY invariants:  they need antifields and they are not superspace \ci{superspace, bagger} invariant.  In equation (\ref{thebasic}) and its component form (\ref{simplestwithoutspin}) a more complete example of an exotic invariant is coupled to a chiral dotted spinor superfield. In 
(\ref{thebasic3}) and its component form 
(\ref{thebasic3comp}), we exhibit the slightly more complicated form that can generate triangle diagrams needed for SUSY anomalies.  In 
(\ref{electron})  there is  an example   in the massless SSM.

\refstepcounter{orange}
{\bf \theorange}.\;
{\bf   The SUSY BRS transformations $\d$, in equation (\ref{Purechiral1}), are nilpotent:}  
We will drop the index $i$ briefly, since for the next few parts we do not need to consider multiple fields. 
Nilpotency means that \be
\d^2 =0.
\ee
This is easy to see.  For example we have 
\be
\d^2 A=  
 \d \y_{  \b} {C}^{  \b} 
+\d \x^{\g \dot \d} \partial_{\g \dot \d} A
- \x^{\g \dot \d} \partial_{\g \dot \d} \d A
\eb
=  
\lt (    
\pa_{ \b \dot \b }  A {\ov C}^{\dot \b}  
+ 
C_{\b}   
F
+ \x^{\g \dot \d} \partial_{\g \dot \d}  \y_{\b  }
\rt ){C}^{  \b} 
- C_{\g} \oC_{\dot \d} \partial^{\g \dot \d} A
- \x^{\g \dot \d} \partial_{\g \dot \d} 
\lt (  \y_{  \b} {C}^{  \b} 
+ \x^{\a \dot \b} \partial_{\a \dot \b} A
\rt)=0
\ee
which follows from simply collecting terms. Note that  the SUSY ghost $C_{\a}$ is Grassmann even, and so: \be
C_{\a} C^{\a}=0
\ee
It is essential that the operator $\d$ is Grassmann odd here, so $A, F, \pa, C$ are all Grassmann even, and $\y, \x,\d$ are all Grassmann odd.

\refstepcounter{orange}
{\bf \theorange}.\;
\la{simplestanomaly}
{\bf  Simplest Example  of SUSY BRS ghost charge one cohomology:}  Lets look at the simplest example for the transformations above.  
\be
\W_1 = \int d^4 x \oF C_{\a};\; {\ov \W}_1 = \int d^4 x F \oC_{\dot \a}
\la{ex1}
\ee
 The second expression ${\ov \W}_1$ is just the complex conjugate of the first ${ \W}_1$.    Both clearly have ghost charge one.  We now will verify explicitly that  ${ \W}_1$ is  in the BRS cohomology space $\cH$.
\be
\cH = \rm \fr{Cocycles \;of\; \d}{Coboundaries \;of\; \d}
\ee

\refstepcounter{orange}
{\bf \theorange}.\;
{\bf  Cocycle:} First we note that $\W_1$ is a cocycle of the $\d$ operator, simply because $C_{\a}$ and  $\x^{\g \dot \d}$ are taken as constant, for rigid SUSY, and we assume that the integral of a total derivative is zero: 
\be
\d \W_1 = \int d^4 x \; \d \oF  \; C_{\a}= \int d^4 x 
\lt ( \pa_{ \g \dot \b }  \oy^{\dot \b} C^{\g}
 + \x^{\g \dot \d} \partial_{\g \dot \d}  \oF \rt )
 C_{\a}=0
\la{ex3}
\ee

\be
\la{Purechiral1}
\framebox{
{$\begin{array}{lll}  
 {\rm Nilpotent}&{\rm WZ}  &{\rm \;Transformations.}
\\d A^i&= & 
  \y^{i}_{  \b} {C}^{  \b} 
+ \x^{\g \dot \d} \partial_{\g \dot \d} A^i
\\
\d {\ov A}_i&= & 
 {\ov \y}_{i  \dot \b} {\ov C}^{ \dot  \b} 
+ \x^{\g \dot \d} \partial_{\g \dot \d} {\ov A}_i
\\
 \d \y_{\a}^i &  =& 
\pa_{ \a \dot \b }  A^{i} {\ov C}^{\dot \b}  
+ 
C_{\a}   
F^i
+ \x^{\g \dot \d} \partial_{\g \dot \d}  \y^{i}_{\a  }
\\
\d
 {\ov \y}_{i \dot \a} &  =& 
\pa_{ \a \dot \a }  {\ov A}_{i} {C}^{\a}  
+ 
{\ov C}_{\dot \a}   
{\ov F}_{i}
+ \x^{\g \dot \d} \partial_{\g \dot \d} 
 {\ov \y}_{i \dot \a} 
\\ 
\d F^i &  =&  \pa_{ \a \dot \b }  \y^{i \a}  {\ov C}^{\dot \b}  
 + \x^{\g \dot \d} \partial_{\g \dot \d}  F^i
\\ 
\d \oF_i &  =&  \pa_{ \a \dot \b }  \oy^{\dot \b}_i  C^{\a}
 + \x^{\g \dot \d} \partial_{\g \dot \d}  \oF_i
\\
\d \x_{\a \dot \b} &  =& - C_{\a} \oC_{\dot \b}
\\
\d C_{\a} &  =& 0
\\
  \d \oC_{ \dot \b} &  =& 0
\\
\end{array}$} }
\ee

\refstepcounter{orange}
{\bf \theorange}.\;
\la{notaboundary1}
{\bf  Verification that (\ref{ex1}) is  not a coboundary:} Next we note that (\ref{ex1})  is  not a coboundary of the  $\d$ operator.  There are only two possible objects that could conceivably give us the object (\ref{ex1}) and neither of them work:
 \be
\cB_1 =  \int d^4 x \;  \y_{\a}, {\ov\cB}_1 =  \int d^4 x \;  \oy_{\dot \a}
\ee
because
\be
\d {\cB}_1  =  \int d^4 x   \lt ( \pa_{ \a \dot \b }  A^{} {\ov C}^{\dot \b}  
+ 
C_{\a}   
F^i
+ \x^{\g \dot \d} \partial_{\g \dot \d}  \y^{}_{\a  }\rt ) 
=  \int d^4 x   \lt ( 
C_{\a}   
F
\rt ) 
\ee
\be
\d {\ov\cB}_1  =  \int d^4 x   \lt (\pa_{ \a \dot \a }  {\ov A} {C}^{\a}  
+ 
{\ov C}_{\dot \a}   
{\ov F}
+ \x^{\g \dot \d} \partial_{\g \dot \d} 
 {\ov \y}_{\dot \a}  \rt ) 
 =  \int d^4 x  
 \lt ({\ov C}_{\dot \a}    {\ov F}
  \rt )  
\ee
 Of the four possible cocycles formed by products  of $C,\oC$ and $F,\oF$ we get two as coboundaries, but the other two in (\ref{ex1}) 
 are cocycles that are not coboundaries, and so they are in the cohomology space.  These are the simplest examples and there are many others described in \ci{dixminram}.  The chief characteristic of them all is that they have unsaturated Lorentz indices, like $\a$ in the above.

\refstepcounter{orange}
{\bf \theorange}.\;
\la{simplestsourceofanom}
{\bf  Simplest version of an exotic invariant:} In order to see that it is possible to generate anomalies corresponding to the cohomology discussed above in (\ref{simplestanomaly}), we need to look at some of the  BRS cohomology of the full Wess-Zumino model.  We introduce the action and notation below in paragraph \ref{actionsec}.  But first let us show what the simplest hint of an exotic invariant  SUSY operator that could generate an anomaly looks like--it is very similar to the expressions (\ref{ex1}):  

\be
\cE_1  = \int d^4 x {\ov \G} C_{\a};\;
{\ov \cE}_{1}  = \int d^4 x   \G \oC_{\dot \a}
\la{simpexotic}
\ee
Here the Zinn-Justin source $\G_i$ is the source for $\d A^i$ as we can see from (\ref{physicaltable}) in paragraph \ref{brspara}, and from the form of the action in paragraph \ref{actionsec}.  The variation of $ {\ov \G}^i$, for the free massless case, is:
\be
\d {\ov \G}
=
- \fr{1}{2} \pa_{ \a \dot \b  }       \pa^{ \a \dot \b  }        { A} 
- \pa_{ \a \dot \b } {\ov Y}^{ \dot \b}    {C}^{\a}   
+ \x^{\g \dot \d} \partial_{\g \dot \d} 
 {\ov \G}
\la{ex5}
\ee
so that we get a zero result $\d \cE_1=0$, 
which is very similar to (\ref{ex3}) when we apply (\ref{ex5}) 
to (\ref{simpexotic}). Moreover if we try to find a boundary to get  (\ref{simpexotic}), the only possible guesses are: 
 \be
\cB_2=  \int d^4 x \;  \Y^{ \a} ,{\ov \cB}_2 =  \int d^4 x \;  \oy^{\dot \a}
\ee
and from (\ref{physicaltable}) in paragraph \ref{brspara},  for the free massless case, we get
\be
\d Y^{ \a}  
=
-
  \pa^{\a \dot \b  }   
{\ov \y}_{  \dot \b}
-
\G
 {C}^{  \a}
+ \x^{\g \dot \d} \partial_{\g \dot \d}  Y^{ \a}
;\;
\d 
{\ov Y}^{ \dot \a} 
= 
-
  \pa^{\b \dot \a  }   
{ \y}_{ \b}
-
{\ov \G}  
 {\ov C}^{\dot  \a}
+ \x^{\g \dot \d} \partial_{\g \dot \d}  
{\ov Y}^{\dot \a} 
\la{oYfree}
\ee
So we find almost exactly the same situation that we noted in paragraph \ref{notaboundary1} above.  From the four cocycles generated from $\G,\ov \G$ multiplied by $C, \oC $, we can generate only two of them using (\ref{oYfree}).  The other two, in (\ref{simpexotic}),
 are in the cohomology space.  These expressions (\ref{simpexotic})
are the simplest examples of exotic invariants.  But they are not very interesting because they do not contain any fields--they only have an antifield and a ghost.

\refstepcounter{orange}
{\bf \theorange}.\;
{\bf The \exinv} \ in paragraph \ref{simplestsourceofanom} and the anomaly expression in paragraph \ref{simplestanomaly}
are valid only for the free and massless theory.  Obviously that theory has no anomalies because it has no Feynman diagrams at all.  But these two simplest examples will be found to be at the origin of the BRS cohomology of these theories also for the massive and interacting theories.  The only difference is that the presence of interactions and masses generate the constraint equations in paragraph \ref{tricky}.  Now we need to go into the details of the massive interacting chiral SUSY Wess Zumino (WZ) model, to see how those arise.

\refstepcounter{orange}
{\bf \theorange}.{\bf \; The Wess Zumino (WZ) Action with masses and interactions} We need this to show how to construct the expressions that can be anomalous in this theory.  
\la{actionsec}
\begin{equation} \cA = -
\int d^{4}x \lt \{
\partial_{\mu} A^a \partial^{\mu} {\ov A}_a
+ \y^{a \a}\pa_{\a \dot \b}  \ov{\y}_a^{\dot \b} - F^a \ov{ F}_a 
\ebp
+
m g_{(ij)}
\lt ( 
A^i F^i
-
\fr{1}{2}\y^{i \b}  
\y^{j}_{ \b}  
\rt )
+
\om \og^{(ij)}
\lt ( \A_i \oF_{1,j} 
-
\fr{1}{2}
\oy_i^{\dot \b}  
\oy_{j,\dot \b}  
\rt )
+ g_{(ijk)} \lt (
F^i A^j A^k -  \y^{i\a} \y^j_{\a} A^l \rt )
\ebp
+ \overline{g}_{abc} \overline{F}_a \overline{ A}_b \overline{ A}_c - \overline{g}^{abc} \overline{\y}_a^{\dot \a} \overline {\y}_{b \dot \a} 
\overline{ A}_c
+ \Lam_a  \lt ( \oC^{\dot \b} \pa_{\a \dot \b} \y^{a \a}+
\x^{\mu} \partial_{\mu} F^a 
\rt )+  \G_a  \lt ( C^{\a} \y^{a}_{\a}+
\x^{\mu} \partial_{\mu} A^a 
\rt )
\ebp
+ Y_a^{\a}
\lt (  \pa_{\a \dot \b}   A^a
\overline{C}^{\dot{\b}}
+ F^a C_{\a}
+ \x^{\mu} \partial_{\mu} \y^a_{\a} \rt ) 
+ \ov\Lam^a  \lt ( C^{\a} \pa_{\a \dot \a} \oy_{a}^{ \dot \a}+
\x^{\mu} \partial_{\mu} \oF_a 
\rt )
+ \ov\G^a  \lt ( \oC^{\dot \a} \oy_{a \dot \a}+
\x^{\mu} \partial_{\mu} \A_a 
\rt )\ebp
+ \oY^{a \dot \a}
\lt (  \pa_{\a \dot \a}   \A_a 
 C^{\a}
+ \oF_a \oC_{\dot a}
+ \x^{\mu} \partial_{\mu} \oy_a^{ \dot\a} \rt ) 
\rt \} + X_{\a \dot \b} C^{\a} \overline{C}^{\dot{\b}}
\ee

\refstepcounter{orange}
{\bf \theorange}.{\bf \; 
The Master equation (Zinn-Justin identity) is:}
\be
\int d^4 x \lt \{
\fr{\d \cA}{\d A^a} \fr{\d \cA}{\d \G_a}
+
\fr{\d \cA}{\d \y^{a\a}} \fr{\d \cA}{\d Y_{a\a}}
+
\fr{\d \cA}{\d F^a} \fr{\d \cA}{\d \Lam_a}
+
\fr{\d \cA}{\d \A_a} \fr{\d \cA}{\d \ov\G^a}
+
\fr{\d \cA}{\d \oy_a^{\dot \a}} \fr{\d \cA}{\d \oY^a_{\dot \a}}
+
\fr{\d \cA}{\d \oF_a} \fr{\d \cA}{\d \ov\Lam^a}
\rt \} + 
\fr{\pa \cA}{\pa X_{\a \dot \b}} 
\fr{\pa \cA}{\pa \x^{\a \dot \b}}   =0
\la{master}
\ee

\refstepcounter{orange}
{\bf \theorange}.{\bf \; 
The nilpotent BRS transformations including antifields, masses and interactions are:}
\la{brspara}
\be
\la{physicaltable}
\vspace{.1in}
\framebox{{$\begin{array}{lll}  
\\ & &{\rm Nilpotent  \;Transformations \;including\;antifields}\\
\d A^i&= & 
\fr{\d {\cal A}}{\d \G_i} 
=  \y^{i}_{  \b} {C}^{  \b} 
+ \x^{\g \dot \d} \partial_{\g \dot \d} A^i
\\
\d {\ov A}_i&= & 
\fr{\d {\cal A}}{\d {\ov \G}^i} 
=  {\ov \y}_{i  \dot \b} {\ov C}^{ \dot  \b} 
+ \x^{\g \dot \d} \partial_{\g \dot \d} {\ov A}_i
\\

\d \y_{\a}^i &  =& 
\fr{\d {\cal A}}{\d {  Y}_i^{   \a} } = 
\pa_{ \a \dot \b }  A^{i} {\ov C}^{\dot \b}  
+ 
C_{\a}   
F^i
+ \x^{\g \dot \d} \partial_{\g \dot \d}  \y^{i}_{\a  }
\\

\d
 {\ov \y}_{i \dot \a} &  =& 
\fr{\d {\cal A}}{\d { {\ov Y}}^{i \dot   \a} } = 
\pa_{ \a \dot \a }  {\ov A}_{i} {C}^{\a}  
+ 
{\ov C}_{\dot \a}   
{\ov F}_{i}
+ \x^{\g \dot \d} \partial_{\g \dot \d} 
 {\ov \y}_{i \dot \a} 
\\
 \d F^i 
&=&
  \pa_{\a \dot \b}   \y^{i \a} {\ov C}^{\dot \b} 
+ \x^{\g \dot \d} \partial_{\g \dot \d}  F^i 
\\
 \d \oF_i 
&=&
  \pa_{\a \dot \b}   \oy_{i}^{\dot \a} { C}^{\b} 
+ \x^{\g \dot \d} \partial_{\g \dot \d}  \oF_i 
\\
\d \G_i 
&= &
 \fr{\d {\cal A}}{\d A^i} 
=
 - \fr{1}{2} \pa_{ \a \dot \b  }       \pa^{ \a \dot \b  }        {\ov  A}_{i} 
  +   m {g}_{iq} F^q  + g_{ijk} ({ A}^{j}{ F}^{k}- {\y}^{j\a}{\y}_{\a}^{k})
\\
&&
-\pa_{ \a \dot \b } Y_{i}^{ \a}    {\ov C}^{\dot \b}   
+ \x^{\g \dot \d} \partial_{\g \dot \d} \G_i
\\
\d {\ov \G}^i 
&= & \fr{\d {\cal A}}{\d {\ov A}_i} 
=
- \fr{1}{2} \pa_{ \a \dot \b  }       \pa^{ \a \dot \b  }        { A}^{i} 
+  m {\ov g}^{ij} {\ov  F}_{ j}  
+  {\ov g}^{ijk}      ({ \A}_{j}{ \oF}_{k}- {\oy}_{j}^{\dot \a}{\oy}_{k \dot \a})
\\
&&
- \pa_{ \a \dot \b } {\ov Y}^{ i \dot \b}    {C}^{\a}   
+ \x^{\g \dot \d} \partial_{\g \dot \d} 
 {\ov \G}^i
\\
\d Y_{i}^{ \a} 
&=&\fr{\d {\cal A}}{\d {  \y}^i_{   \a}} 
= 
-
  \pa^{\a \dot \b  }   
{\ov \y}_{i   \dot \b}
+  m {g}_{iq}   
\y^{q \a} 
\\
&&
 +
2 g_{ijk}  \y^{j \a} A^k    
-
\G_i  
 {C}^{  \a}
+ \x^{\g \dot \d} \partial_{\g \dot \d}  Y_{i}^{ \a}
\\
\d 
{\ov Y}^{i \dot \a} 
&=&\fr{\d {\cal A}}{\d {\ov \y}_i^{ \dot \a} 
} 
= 
-
  \pa^{\b \dot \a  }   
{ \y}^i_{ \b}
+  m {\ov g}^{ik}   
{\ov \y}_{k}^{\dot  \a} 
\\
&&
+
2 {\ov g}^{ijk} {\ov \y}_{j}^{\dot  \a} 
{\ov A}_k  
-
{\ov \G}^i  
 {\ov C}^{\dot  \a}
+ \x^{\g \dot \d} \partial_{\g \dot \d}  
{\ov Y}^{i \dot \a} 
\\
 \d \Lam_i  
&=&  \fr{\d {\cal A}}{\d {F}^i } =
\oF_i + m g_{ib}  A^b+  g_{ibc}  A^b A^c
+ Y_i^{\a}
  C_{\a}+ 
\x^{\g \dot \d} \partial_{\g \dot \d}   \Lam_i  
\\
 \d {\ov \Lam}^i  
&=&  \fr{\d {\cal A}}{\d {\ov F}_i } =
F^i +m \og^{ij}   \A_j + \og^{ijk}  \A_j \A_k
+ \oY^{i \dot \a}
  \oC_{\dot\a}+ 
\x^{\g \dot \d} \partial_{\g \dot \d}   {\ov \Lam}^i   
\\

\d \x_{\a \dot \b} 
&=& \fr{\pa {\cal A}}{\pa { X}^{\a \dot    \b}} 
=
 -   C_{\a} {\ov C}_{\dot \b}
\\
\d X_{\a \dot \b} 
&=& \fr{\pa {\cal A}}{\pa { \x}^{\a \dot    \b}} 
= 
\int d^4 x \; \X_{\a \dot \b} 
\\
\d C_{\a}
&=&
0
\\
\d  {\ov C}_{\dot \b}
&=&
0
\\
\end{array}$}} 
\ee

\refstepcounter{orange}
{\bf \theorange}.\;
{\bf The full nilpotent $\d$:}  Now that we have introduced  the WZ action and Zinn-Justin's sources, we can discuss the BRS cohomology of the full nilpotent $\d$ above.

\refstepcounter{orange}
{\bf \theorange}.\;
\la{simplestexoticpair} 
{\bf   Simplest Lorentz invariant exotic pair:} 
Consider the $\d$ that corresponds to the massless free WZ action with  antifields.  These are summarized in equations   (\ref{brspara}), if we make those massless and free by setting $m g_{ij} \ra 0, g_{ijk } \ra 0$.  Then the following expressions $[\cE,\W]$  are an exotic pair  with ghost charges zero and one:
\be
\cE= \int d^4 x d^2 \oq \lt \{
\lt ( {\widehat \f}^{\a}  {\widehat {\ov \Lam}} 
\rt ) C_{\a}  \rt \}+
 \int d^4 x d^4 \q \lt \{ {\widehat \f}^{ \a}\q_{\a}   {\widehat A}    \rt \}  \in \cH
\la{thebasic}
\ee
\be
\W= \int d^4 x d^2 \oq \lt \{
 {\widehat \f}^{\a}    {\widehat \A}^2 C_{\a}   \rt \}  \in \cH
\la{thebasicghostone}
\ee
It is crucial that each member of the exotic pair has the same mass dimension, and also has the same quantum number for things like lepton number charge, etc. There are no quantum numbers in this case, but these both have the same mass dimension
 ${\rm Dim}[\cE]={\rm Dim}[\W]=  m^{-1}$, where we take 
\[
{\rm Dim} \lt [d x^{\m}\rt ]= m^{-1}, {\rm Dim} \lt[\fr{\pa}{\pa x^{\m}}\rt]= m^{1},{\rm Dim}[\q_{\a}]= {\rm Dim} [C_{\a}]= m^{-\fr{1}{2}}, {\rm Dim} [\Lam]= m^{2}  \] \be  {\rm Dim}\lt [d \q^{\a} \rt ]={\rm Dim}\lt [\fr{\pa}{\pa \q_{\a}}\rt ]=   m^{+\fr{1}{2}},{\rm Dim} [A]= m^{1}, 
{\rm Dim} [\f]= m^{\fr{1}{2}}, 
{\rm Dim} [\y]= m^{\fr{3}{2}}, 
\la{dimension}
\ee
This ghost charge one invariant (\ref{thebasicghostone}) is rather straightforward and we discuss it below in paragraph \ref{tricky}.  
 The expression 
 in (\ref{thebasic}) is shown and discussed  in components  below in paragraph \ref{introsources}.
In the above ${\widehat \f}^{\a}$ is a source, rather than a dynamic field.  It is simply an   undotted antichiral  spinor superfield source.  It satisfies the antichiral constraint:
\be
{\widehat \f}^{\a}= {  \f}^{\a}+\oq_{\dot \b} W^{\a \dot \b}+\fr{1}{2} \oq^2 {\F}^{\a}; 
D_{\a} {\widehat \f}_{\b}=0
\la{sourcephi}\ee

\refstepcounter{orange}
{\bf \theorange}.\;
{\bf  Dotted Chiral Superfields ${\widehat {\ov\f}}^{\dot \a}$ in SUSY:}
The antichiral superfield ${\widehat \f}^{ \a}$ transforms exactly like the antichiral superfield
${\widehat \A}$, except that it has an index $\a$ that is carried along inertly:

\be
\d {\f}_{}^{ \a}=
{W}_{}^{ \a \dot \b} \oC_{\dot \b}
+ \x \cdot \pa {\f}_{}^{ \a}
;\; \d {W}_{}^{ \a \dot \b}=
\pa^{\g \dot \b }  {\f}_{}^{ \a } C_{\g}
+ {\F}_{}^{ \a} \oC^{\dot \b}
+ \x \cdot \pa \; {W}_{}^{ \a \dot \b}
;\;\d {\F}_{}^{ \a}=
\pa_{\g \dot \b }  {W}_{}^{\a \dot \b } C^{\g}
+ \x \cdot \pa \; {\F}_{}^{ \a}
\ee

\refstepcounter{orange}
{\bf \theorange}.\;
{\bf The chiral superfield  ${\widehat {  \Lam}}$} is the chiral scalar superfield used in the action in paragraph \ref{actionsec}, as a Zinn-Justin source  for the variations of the physical field  ${\widehat A}$. The superfield ${\widehat {\ov \Lam}}$ is its complex conjugate.  It is Grassmann odd and it satisfies the antichiral constraint, so it has the component expansion:
\be
{\widehat {\ov \Lam}} = { \ov \Lam} +\oq_{\dot \b} \oY^{ \dot \b}+\fr{1}{2} \oq^2 {\ov \G}; D_{\a} {\widehat {\ov \Lam}} 
\la{thebasiclam}
\ee

\refstepcounter{orange}
{\bf \theorange}.\;
{\bf  
The complex two-component spinor}  $C_{\a}$ is the constant Grassmann even spinor ghost of SUSY, and $\q^{\a} $ is the constant Grassmann odd spinor coordinate  of superspace. The basic chiral matter superfield here is:
\be
{\widehat A} = { A} +\q_{\a} \y^{ \a}+\fr{1}{2} \q^2 {F}; \ovD_{\dot\b} {\widehat A}=0
\la{thebasicA}
\ee
This notation is used  to formulate the WZ action and the BRS transformations in components.

\refstepcounter{orange}
{\bf \theorange}.{\bf \;Introduction of sources}
\la{introsources}
Using ${\widehat \f}^{\a}$  as if it were analogous to ${\widehat \A}$ we get the following in place of (\ref{simplestsourceofanom}).  This is the component version of the superspace expression in (\ref{thebasic}) above:
\be
\cE  = \int d^4 x \lt \{
\lt ( \f^{\a} \ov \G  
+2 W^{\a \dot \b} \ov Y_{\dot \b}  
+\F^{\a} \ov \Lam \rt ) C_{\a}  
-\F^{\a} \y_{\a}
 -W^{\a \dot \b} \pa_{\a \dot \b}A \rt \} \in \cH 
\la{simplestwithoutspin}
\ee
Let us verify that this is invariant first, and see why. Consider the first part first:
\be
\d \int d^4 x \lt \{
\f^{\a} \ov \G C_{\a} 
+ 2 W^{\a \dot \b} \ov Y_{\dot \b} C_{\a}  
+\F^{\a} \ov \Lam C_{\a}  \rt \}
\ee
This is zero for the SUSY  transformations plus the following for the equation of motion parts:
\be
  \int d^4 x \lt \{
\f^{\a} \Box A C_{\a} 
+ 2 W^{\a \dot \b} \lt ( \pa_{\g \dot \b} \y^{\g}  \rt ) C_{\a}  
+\F^{\a} F C_{\a}  \rt \}
\la{firstpartfromantifields}
\ee
Now  consider the second part. It has only SUSY variations, since there are no antifields present in it.  But it is not SUSY invariant: 
\be
\d  \int d^4 x \lt \{
\F^{\a} \y_{\a}
 +W^{\a \dot \b} \pa_{\a \dot \b}A \rt \}  
=  \int d^4 x \lt \{
\d \F_{\a} \y^{\a}
 - \F_{\a} \d \y^{\a}
 +\d W^{\a \dot \b} \pa_{\a \dot \b} A   
 +W^{\a \dot \b} \pa_{\a \dot \b}\d A \rt \}  
\ee
which is 
\be
=  \int d^4 x \lt \{
 \pa_{\g \dot \b}  W^{\a \dot \b} C^{\g}
 \y^{\a}
 - \F^{\a} 
\lt (  \pa_{\a \dot \b}   A
\overline{C}^{\dot{\b}}
+ F C_{\a}
\rt ) 
 +\lt (
\pa_{\g \dot \b} 
\f_{\a} C^{\g}
+ \F_{\a}\oC_{\dot \b}
\rt )
  \pa_{\a \dot \b} A   
 +W^{\a \dot \b} \pa_{\a \dot \b}
C^{\g} \y_{\g}
 \rt \}  =
\eb 
  \int d^4 x \lt \{
 +
\pa_{\g \dot \b} 
\f_{\a} C^{\g} \pa_{\a \dot \b} A   
- \pa_{\g \dot \b}  W^{\a \dot \b} C^{\g}
 \y^{\a} +W^{\a \dot \b} \pa_{\a \dot \b}
C^{\g} \y_{\g}
- \F^{\a}  F C_{\a}
 \rt \}  
\la{secondpartfromSUSY h}
\ee

After integration by parts, this exactly cancels (\ref{firstpartfromantifields}).

\refstepcounter{orange}
{\bf \theorange}.\;
\la{summary}
{\bf  In summary,}
by taking the antichiral undotted spinor superfield ${\widehat \f}^{\a}$ from  (\ref{sourcephi})  and contracting it with $C_{\a} $,   and coupling that  to the antichiral complex conjugate antifield multiplet   ${\widehat {\ov \Lam}} $ from 
(\ref{thebasiclam}), and integrating that combination over $\int d^4 x d^2 \oq$, and then adding the SUSY non-invariant terms as in line (\ref{simplestwithoutspin}), we have the following results:
\bitem
\item

The expression $\lt ( \f^{\a} \ov \G  
+ 2 W^{\a \dot \b} \ov Y_{\dot \b}  
+\F^{\a} \ov \Lam \rt )$ is, to start with, just a normal antichiral SUSY invariant $ \int d^4 x d^2 \oq \lt \{
\lt ( {\widehat \f}^{\a}  {\widehat {\ov \Lam}}
\rt )   \rt \}$  
\item
However, the expression $\lt ( \f^{\a} \ov \G  
+2 W^{\a \dot \b} \ov Y_{\dot \b}  
+\F^{\a} \ov \Lam \rt )$ also gets a variation from 
the equation of motion terms in the antifield BRS variations, and the result of that is (\ref{firstpartfromantifields}).
\item
Then we consider the SUSY variation of the field dependent parts $-\F^{\a} \y_{\a}
 -W^{\a \dot \b} \pa_{\a \dot \b}A$ of (\ref{simplestwithoutspin}).  These parts are not SUSY  invariant, but these parts are  arranged so that their variations (\ref{secondpartfromSUSY h})
 just cancel 
(\ref{firstpartfromantifields}).
\item So the combination $\cE $ 
formed by the sum in  
(\ref{simplestwithoutspin})
is actually an invariant of the two parts of the BRS variation. 
But it is not a SUSY invariant. It is only a BRS invariant, and they are not the same.
\item The proof that this set of facts is possible arises from the spectral sequence analysis, as we show in paragraphs \ref{specseqsumm} to \ref{trickysum} below.
\item
The constraint equations are also shown in paragraphs \ref{specseqsumm} to \ref{trickysum} below. They are needed for the next example, which is in paragraph \ref{simplestexoticpair}.
\eitem

\refstepcounter{orange}
{\bf \theorange}.\;
{\bf  We have shown} that $\cE $  is a cocycle of our $\d$.  
  In order to see that it is in the cohomology space,  we need to examine the possible coboundaries. We need an object that has ghost charge minus one, and all indices absorbed, and it must have the right dimension, and the right kinds of fields, of course.  What can we make? The only available expression here is:
\be
\cB  = \int d^4 x \lt \{
  \f^{\a}  Y_{\a}   
  \rt \}  
\ee
It yields 
\be
\d \cB  = \int d^4 x \lt \{
 \d \f^{\a}  Y_{\a}   
+  \f^{\a} \d Y_{\a}   
  \rt \}  
 = \int 
d^4 x \lt \{
W_{\a \dot \b} \oC^{\dot \b}  Y_{\a}   
+  \f^{\a}
\lt (
\s^{\m}_{\a \dot \b}
\pa_{\m} \oy^{\dot \b} 
+ \pa_{\a \dot \b}\Lam \oC^{\dot \b}+
\G C_{\a} \rt ) 
  \rt \}  
\ee

\refstepcounter{orange}
{\bf \theorange}.\;
{\bf  Compare} this with  (\ref{simplestwithoutspin}):
\be
\cE  = \int d^4 x \lt \{
\lt ( \f^{\a} \ov \G  
+ W^{\a \dot \b} \ov Y_{\dot \b}  
+\F^{\a} \ov \Lam \rt ) C_{\a}  
-\F^{\a} \y_{\a}
 -W^{\a \dot \b} \pa_{\a \dot \b}A \rt \} \in \cH 
 \ee

The terms $\cE $ and $\d \cB$ are very far from matching, though some terms are a bit similar. 
The failure to match is reminiscent of the similar, but simpler,  cases in sections \ref{simplestanomaly} and
\ref{simplestsourceofanom}.  This is the magic of the spectral sequence.  It guarantees this kind of result.

\refstepcounter{orange}
{\bf \theorange}.{\bf The next exotic pair}
\la{nextosimplestexoticpair}
 Now we turn to the next to simplest  exotic pair.  This is the simplest exotic pair that can occur in a massive or interacting theory.  It also gives rise to the simplest possible anomaly calculation.
Now we do need to consider multiple fields, as will become clear.  So we will include the index (a,b,c or i,j,k) on the field.  
So here is the  simplest form of an exotic pair in a massive or interacting theory:
\be
\cE = \int d^4 x d^2 \oq \lt \{
\lt ( {\widehat \f}^{\a}  \ov z^j_i {\widehat {\ov \Lam}}^i \A_j
\rt ) C_{\a}  \rt \}+
 \int d^4 x d^4 \q \lt \{ {\widehat \f}^{ \a}\q_{\a}  \ov z^j_i   {\widehat A}^i  \A_j \rt \}  
\la{thebasic3}
\ee
\be
\W= \int d^4 x d^2 \oq \lt \{
 {\widehat \f}^{\a} \ov w^{ijk}   {\widehat \A}_i  {\widehat \A}_j {\widehat \A}_k  C_{\a}   \rt \}  \in \cH
\la{thebasicghostone3}
\ee
This pair $[\cE ,\W]$  both have dimension $m^0$ using our dimensions in equation (\ref{dimension}). Also they need to have indices now, since the constraint equations are applicable, and here we can consider the massive and interacting theories with $m g_{ij}\neq 0, g_{ijk } \neq 0$, instead of the free massless theory considered in  paragaph \ref{simplestexoticpair}.   Including the mass and interactions means that one must also satisfy the constraints in paragraph \ref{tricky}. 
The next paragraph discusses those constraints.

\refstepcounter{orange}
{\bf \theorange}.\;
\la{anomconstrainteqs}
{\bf  Anomaly constraint equations:} A crucial feature of the BRS cohomology of chiral SUSY is that  the exotic invariants and the possible anomalies of SUSY are all constrained by certain equations.  For the simplest massless but interacting cases, as in paragraph \ref{nextosimplestexoticpair}, the exotic invariants and their anomalies  can be constructed from expressions of the following form:
\be
\cE'=  z^i_j\oy_{i \dot \a}  A^j ,
 \W' = w_{ijk} A^{i}  A^{j} A^{k} \oC_{ \dot \a}
\la{examples}
\ee 
$\cE'$ can be used to construct the ghost number zero exotic invariant, and  $\W'$  can be used to construct  the corresponding ghost number one anomaly\footnote{The primes on $\cE'$   and  $\W'$  indicate that the objects are in the spectral sequence space $E_{\infty} $,  which is isomorphic to, but much simpler than, and very different from, the  cohomology space $\cH$. The isomorphism $E_{\infty} \approx \cH$, and its trickiness, are discussed in paragraph \ref{tricky}.}.  Here $A^i$ and $\oy_{i \dot \a}$ are the scalar and complex conjugate spinor of a Wess Zumino theory, $i$ is an index labelling the different fields, and $\oC_{\dot \a}$ is the complex conjugate of the constant SUSY spinor ghost of that theory. We show below how to transform  (\ref{examples})  into expressions in the quantum field theory, which can then be used to generate triangle diagrams to calculate the SUSY anomalies.  
As we show below in paragraph \ref{d2}, using spectral sequence methods, the coefficient tensors are constrained by the BRS cohomology to satisfy the following constraints:
\be
  z^m_{(i}    g_{jk)m}  =0
;\;    \og^{ijm} w_{ijn}    =0
\ee
where   the superpotential $\cP$, and its complex conjugate $\ov \cP$, are defined by:
\be
\cP= g_{ijk} A^i A^j A^k;\;
{\ov \cP}= \og^{ijk} \A_i\A_j\A_l
\ee
 The simplest known solutions, at present, to these constraints, are in the supersymmetric standard model (SSM), as shown below in paragraph \ref{ssmexample}.

\refstepcounter{orange}
{\bf \theorange}.{\bf Feynman Triangle Diagramsy:}
\la{feyntri}
The expansion of the second part is the basis of our triangle diagrams:
\be
 \int d^4 x d^4 \q \lt \{ {\widehat \f}^{ \a}\q_{\a}  z^j_i   {\widehat A}^i  \A_j    \rt \}  
\ra 
\int d^4 x z^j_i \lt \{  \F^{\a} \A_j \y^i_{\a}+W^{\a \dot \d} \lt ( 
\oy_{j\dot \d}\y^i_{\a} 
+  \A_j  \pa_{\a \dot \b}A^i\rt )
+ \f^{\a}\lt ( 
\oF_j \y^i_{\a}
+2 \oy_j^{\dot \b} \pa_{\a \dot \b}A^i\rt )
   \rt \}  
\la{thebasic3comp}
\ee

\be
\d {\cal G}_1[\f, A, \oy] \approx  
\lt \{ \int d^4 x \pa \A C \fr{\d }{\d \oy} + \cdots \rt \}
{\cal G}_1[\f, A, \oy] 
\approx \int d^4 x \lt \{ \cdots
   \f_{\a} C^{\a} \Box A^i  {\ov A}_{j }  + \cdots
\rt \} 
\ee

\refstepcounter{orange}
{\bf \theorange}.\;
{\bf  If needed for a calculation, we can add a finite local renormalization ghost charge zero term of the form}

\be
\cR = -\int d^4 x \lt \{ \cdots
+  \f_{\a} C^{\a} \ov \G^i {\ov A}_{j }  
\rt \} 
\ee
and  noting that 
\be
\d \ov \G^i = \Box A^i +  (\og \A \oF)^i + \cdots 
\ee
 we could get
\be
\d \lt ( \cG + \cR \rt )= \int d^4 x \lt \{ \cdots
+ e_i^j \f_{\a} C^{\a} (\og \A \oF)^i  {\ov A}_{j }  + \cdots
\rt \} 
\ee
so that we could generate a term that looks like it is in the cohomology space. Here we have not tried to be careful about the factors of $m$, integration by parts, shifting linearly divergent integrals, matching indices, etc.    The only purpose here is to show one way that the anomaly terms without any derivatives can appear from the linearly divergent integral, and then part of it could be used to convert the term as follows 
\be
\int d^4 x \lt \{ \f^{\a} C_{\a}   \A \Box A + \f^{\a} C_{\a}   \A  \d \oG+ \cdots
\rt \}
\Ra 
\int d^4 x \lt \{ \f^{\a} C_{\a}   \A \lt (  \A \oF +  \oy \cdot \oy  \rt )   
+ \cdots\rt \}
\ee
and so we could generate part of the anomaly in canonical form, using
 \be
\d \oG= \Box A +\og  \lt ( \A \oF + \oy \cdot \oy \rt )   
\ee

\refstepcounter{orange}
{\bf \theorange}\;
\la{notationforssm}
{\bf Our notation for the Superstandard Model is as follows:}

\la{ssmnot}
\vspace{.2cm}
\begin{tabular}{|c|c|c
|c|c|c
|c|c|c|}
\hline
\multicolumn{8}
{|c|}{ \bf Superstandard Model, Left ${\cal L}$
Fields}
\\
\hline
{\rm Field} & Y 
& {\rm SU(3)} 
& {\rm SU(2)} 
& {\rm F} 
& {\rm B} 
& {\rm L} 
& {\rm D} 
\\
\hline
$ L^{pi} $& -1 
& 1 & 2 
& 3
& 0
& 1
& 1
\\
\hline
$ Q^{cpi} $ & $\fr{1}{3}$ 
& 
3 &
2 &
3 &
 $\fr{1}{3}$
& 0
& 1
\\\hline
$J$
& 0 
& 1
& 1
& 1
& 0
& 0
& 1
\\
\hline
\multicolumn{8}
{|c|}{ \bf Superstandard Model, Right 
${\cal R}$
Fields}
\\
\hline
$P^p$ & 2 
& 1
& 1
& 3
& 0
& -1
& 1
\\
\hline
$R^p$ & 0 
& 1
& 1
& 3
& 0
& -1
& 1
\\

\hline
$T_c^p$ & $-\fr{4}{3}$ 
& ${\ov 3}$ &
1 &
3 &
 $-\fr{1}{3}$
& 0
& 1
\\
\hline
$B_c^p$ & $\fr{2}{3}$ 
& ${\ov 3}$ 
&
1 &
3 &
 $- \fr{1}{3}$
& 0
& 1
\\
\hline
$H^i$ 
& -1 
& 1
& 2
& 1
& 0
& 0
& 1
\\
\hline
$K^i$ 
& 1 
& 1
& 2
& 1
& 0
& 0
& 1
\\
\hline
\end{tabular}
\begin{tabular}{|c|c|c
|c|c|c
|c|c|c}
\hline
\multicolumn{8}{|c|}{\bf  Yukawa for Superstandard Model}
\\
\hline
\multicolumn{8}{|c|}{ ${\cal Y} \approx 
g_{ijk}{\cal L}^i {\cal R}^j {\cal R}^k
- g' m^2 J
$}
\\
\hline
\multicolumn{8}{|c|}{$
{\cal Y} =
g' \e_{ij} H^i K^j J
- g' m^2 J
$}
\\
\multicolumn{8}{|c|}{$
+
p_{pq} \e_{ij} L^{p i} H^j P^q
+
r_{pq} \e_{ij} L^{p i} K^j R^q
$}\\
\multicolumn{8}{|c|}{$
+
t_{pq} \e_{ij} Q^{c p i} K^j T_c^q
+
b_{pq} \e_{ij} Q^{c p i} H^j B_c^q
$}
\\
\hline
\end{tabular}
\\

\refstepcounter{orange}
{\bf \theorange}.\;
\la{ssmexample}.
{\bf  An example of an exotic pair   in  the Superstandard Model}    
Here is an example of a \exinv\   that has the quantum numbers of the electron flavour triplet $e,\m,\t$.  We put the mass to zero here, and the notation is summarized in in paragraph \ref{notationforssm} \ci{susyconf}.   
\[
\cE'= (C \x^2 C) z_p \f^{\a} \lt (g' \oK_i \y_{L\a}^{ i p } - \ov p^{pq}
 \oP_q \y_{J a}\rt  )
\stackrel{ d_2 }{\lra}
(C \x^2 C)  \f^{\a} z_p g' \oK_i \lt (\ov p^{pq} \oH^i {\ov P}_{q}C_{\a}+
\ov r^{pq} \oK^i \ov N_{q}C_{\a}\rt)
\]
\be
-(C \x^2 C) \f^{\a} z_p  \ov p^{pq}
 \oP_q g'
\oH^i   \oK_i C_{\a}=0\;{\rm where \;we\; note\;that}\; \oK^i\oK_i=0
\la{electron}
\ee
and
\be
\W'= (C \x^2 C)  w^q   
 \oP_q \oJ  \oJ \f^{\a} C_{\a}\ra \W  
= 
\eb
\int d^4 x C_{\a} \lt \{   w^q   
 \oP_q \oJ  \oF_J \f^{\a}  + {\rm \;many \;terms\;  with\; two \;\oJ\; type\; terms}\rt \}
\la{JJanom}
\ee
These both satisfy the anomaly constraint equations in paragraph \ref{tricky}, once those are translated into this notation. 
Now we can take part of the above as follows:
\be
(C \x^2 C)g' z_p \f^{\a}  g' \oK_i \y_{L}^{ i p } \ra \int d^4 x g' z_p
  \lt \{ \f^{\a} \oy^{\dot \b}_{K i} \pa_{\a \dot\b} L^{ i p }
+ \cdots \rt \}
\ee
and illustrate a Feyman triangle diagram constructed from it.

\refstepcounter{orange}
{\bf \theorange}.{\bf Here is the result from the part 
$
g' z_p \f^{\a} \oy^{\dot \b}_{K i} \pa_{\a \dot\b}L^{ i p }$:}

\be
  {\cal G}_1[\f, L, \oy_{K}] = \int d^4 l   d^4 p d^4 q d^4 k\; \d^4(k+p+q)
 \fr{\f (k) \cdot(l+q)\cdot \ov {(l-p)}
 \cdot  (l) \cdot {\ov \y}_{K i}(q) } { (l-p)^2  
 (l+q)^2  l^2    }L^{ i n } (p)
  g' z_t   \ov r^{rt}  p_{r n}
\la{feynint}
\ee

corresponding to the Feynman diagram:

\begin{picture}(250,150)
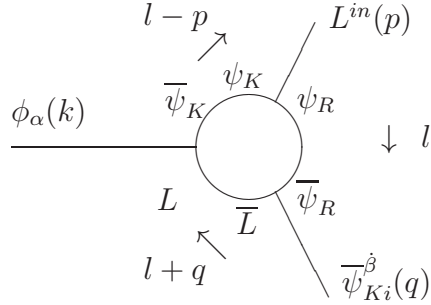
\put(100,02){Feynman Diagram for ${\cal G}_1[\f, L, \oy_{K}] ; k+p+q=0$ }
%
%
%
%
\put(210,92){\line(1,2){15}}
\put(210,58){\line(1,-2){20}}
\put(180,75){\line(-1,0){70}}
\put(200,75){\circle{40}}
%
\put(230,120){$L^{in}(p)$}
\put(235,20){${\ov \y}_{K i}^{ \dot \b}(q)$}
\put(110,85){{$\f_{\a}(k)$}}
\put(168,89){${\ov \y}_{K }$}
\put(190,99){${\y_K }$}
\put(218,90){${\y_R}$}
\put(218,52){${\ov \y}_R$}
\put(195,42){${\ov L}$}
\put(165,51){$L$ }
%
\put(180,110){$\nearrow$}
\put(160,121){$l-p$}
\put(250,75){$\downarrow$}
\put(265,75){$l$}
\put(180,35){$\nwarrow$}
\put(160,25){$l+q$}
%
\end{picture} 

\refstepcounter{orange}
{\bf \theorange}.{\bf \; 
Informal  remarks about the calculation of the supersymmetry anomaly  diagrams:} 
For the massless case we can see that there is a possible anomaly of the form (\ref{JJanom}) but it is clear that no variation of $\cG$ can produce the relevant expression, because terms with 2 J cannot be produced here.

Let us note the following
\ben
\item
The integral (\ref{feynint}) 
is linearly divergent, so it is ripe for the various things that happen when one tries to calculate an anomaly.
\item
However it is clear that no variation  $\d \cG$, and no variation of any possible local counterterm $\d \cal C$,  can produce the relevant anomaly. 

\item
So although there are plenty of exotic pairs in the massless SSM, it looks doubtful that a SUSY anomaly occurs there. 
\item
The electron type exotic invariant in (\ref{electron}), ceases to be an invariant, if one adds mass to the SSM.   The basic result is that, starting with (\ref{electron}) and taking $m\neq 0$ in the superpotential in paragraph \ref{ssmnot}, one finds that
\be
d_2 {\ov \cE}'= 
 (\oC \x^2 \oC) \ov z^p \ov \f^{\dot \a} \oC_{\dot \a} 
  m^2 \og'   
p_{pq}       P^q \neq 0
\ee
This reflects an important change in the Yukawa superpotential when we add mass. The natural conclusion is that these exotic invariants do not exist in the spontaneously broken SSM, although they do exist in the massless SSM.
\item
So that means that there is probably a need to go beyond the SSM, or at least to more complicated expressions in the SSM, to search for exotic pairs that do yield SUSY anomalies. There might also be useful models that are far simpler than the SSM. 
\een

\refstepcounter{orange}
{\bf \theorange}.{\bf The Spectral Sequence}
\la{specseq}
We will now proceed to explain how the spectral sequence predicts 
the existence of these \exinv s, and the constraint equations.
The following parts are based on the methods 
\ci{dixspecseq} as used in \ci{holes,  dixminram, dixmin}.  
The treatment here is very brief, and we only examine a tiny part of the solution.  The purpose of this part is to just be able to see where the \exinv s and the constraint equations come from, and how to use them in practice. 

\refstepcounter{orange}
{\bf \theorange}. {\bf  Grading:} The spectral sequence for an operator is entirely generated by the  grading we choose. Here we choose:
\be
N= N_{\rm all\; fields\;and\;sources}
+
N_{\rm C} 
+
N_{\rm \oC} 
+ 
2 N_{\x}
\la{grading}
\ee
where
\be
N_{\rm all\; fields\;and\;sources} = N_A +N_{\G} +N_\A +N_{\ov \G} + {\rm etc.}
\la{grading2}
\ee
Then we get, for the massless case,
\be  \d= \sum_{r=0}^2 \d_r \; {\rm where}
;\;  [ N , \d_r]=r\d_r
\ee and
\be
\d_0 = \int d^4 x \lt \{ \Box \A \fr{\d}{\d \G}
+\Box A \fr{\d}{\d \oG} 
+ \s^{\m}_{\a \dot \b}
\pa_{\m} \oy^{\dot \b} \fr{\d}{\d Y_{\a}}
+ \s^{\m}_{\a \dot \b}
\pa_{\m} \y^{\a} \fr{\d}{\d \oY_{\dot \b}}
+  \oF \fr{\d}{\d \Lam}
+F \fr{\d}{\d \ov \Lam}\rt \}
+ C   \s^{\m}  \overline{C}   \x^{\mu \dag}  
\la{deltazero}
\ee

\be
\d_1 = \int d^4 x \lt \{ C \y  \fr{\d}{\d A}
+\oC \oy \fr{\d}{\d \A}
+
\lt ( \pa_{\a \dot \b}
   A^a \oC^{\dot \b} + F^a C_{\a} \rt )\fr{\d}{\d \y_{\a}}
 +
\lt ( \pa_{\a \dot \b}
   \A C^{\a} + F^a C_{\a} \rt )
\fr{\d}{\d \oy_{\dot \b}}
\la{thisone2}
\ebp+\lt ( \oC^{\dot \b} \pa_{\a \dot \b} \y^{a \a} 
\rt )   \fr{\d}{\d F^a}
+\lt ( C^{\a} \pa_{\a \dot \a} \oy_{a}^{ \dot \a} 
\rt ) \fr{\d}{\d \oF_a}
+ \lt (Y_a^{\a} C_{\a} +  g_{abc}  A^b A^c   \rt ) \fr{\d}{\d \Lam_a}
+\lt ( \oY^{a \dot\a} \oC_{\dot \a}+ \og^{abc}  \A_b \A_c   \rt )  \fr{\d}{\d \ov \Lam^a}
\la{thisone1}
\ebp 
+\lt ( \pa_{\a \dot \b}\Lam \oC^{\dot \b}+
\G C_{\a} +  g A \y_{\a} \rt ) \fr{\d}{\d Y_{\a}}
+ \lt ( \pa_{\a \dot \a} \ov\Lam C^{\a}+ \ov \G \oC_{\dot \a} + \og \A \oy_{\dot \a}
\rt ) \fr{\d}{\d \oY_{\dot\a}}
+\ebp+   g\lt ( \y^{\a} \y_{\a} +  AF\rt ) \fr{\d}{\d \G}
+ \og\lt ( \oy \cdot\oy +  \A \oF\rt ) \fr{\d}{\d \oG}
\rt \}
\ee

\be
\d_2 =  \x^{\mu} \partial_{\mu}  
\ee

Here we treat ${\widehat \f_{\a}}$ for present purposes as though it were another ${\widehat \A}$ field but with an index $\a$ and no field index $i$.  Also we ignore any action or interaction terms for it, for present purposes.

\refstepcounter{orange}
{\bf \theorange}.{\bf \;Form of $E_1$}
For present purposes we can write
\be
\d_0 = \d_{0, \rm Fields} + \d_{0, \rm Structure}   
\ee As discussed in \ci{dixspecseq}, we start with the operator $\d_0$.  
We easily get the following form, where we show the explicit form in terms of $\x$ and $C, \oC$:  
\[
E_1=M_0 +
 \S_{n=1}^{\infty} P_n  C_{\a_1} \cdots C_{\a_n}
 +\S_{n=1}^{\infty} \oP_n   \oC_{\dot \a_1} \cdots \oC_{\dot \a_n}
+ \S_{n=0}^{\infty} Q_n  (\x C)_{\dot \a}   C_{\a_1} \cdots C_{\a_n}
\]\be
+\S_{n=0}^{\infty} \oQ_n (\x \oC)_{ \a}   \oC_{\dot \a_1} \cdots \oC_{\dot \a_n}
+ (C \x \oC) R_0 
+ \S_{n=0}^{\infty}(C \x^2 C) T_n C_{\a_1} \cdots C_{\a_n}
+\S_{n=0}^{\infty}(\oC \x^2 \oC) \oT_n \oC_{\dot \a_1} \cdots \oC_{\dot \a_n}
\la{structure}
\ee

\refstepcounter{orange}
{\bf \theorange}.\;
{\bf  In the above,}   $M_0$ and $R_0$ are real,  and $P_n,Q_n$ and $T_n$ are complex. These coefficients are all functions of the variables $A^a,\y^a_{\a}$, their complex conjugates $\A_a,\oy_{a\dot\a}$and the derivative operator $\pa_{\a \dot \b}$. But at the stage of $E_1= \ker \d_0 \cap  \ker  \d_0^{\dag}$,  the auxiliary variables $F^a,\oF^a $, and the antifield variables
 $\Lam_a, Y_{a\a}, \G_a$ and their complex conjugates 
$\ov \Lam^a, \oY^a_{\dot\a}, \ov \G^a$, and all their derivatives,  have all been eliminated from the converging set of spaces $E_r$.  The way this works is very easy.  We also will assume that  the $\f,W,\F$ are also reduced simply to the underived $\f_{\a}$.    Much more of this kind of analysis can be found in the references \ci{holes, holescommun,dixminram, dixspecseq,dixmin}. 

\refstepcounter{orange}
{\bf \theorange}.\;
\la{specseqsumm}
{\bf  We will restrict our analysis} here to the following special parts of $E_{\rm 1,special}\subset E_1$ and the subspaces $E_{\rm r,special}\subset E_r$ for $r >1$
\be
E_{\rm 1,special}=
  (C \x^2 C) \f_{ \a} T_a[\A] \y^{a \a}  + (C \x^2 C) \f_{ \a}
T [\A]
 C^{ \a} 
\la{tinybit}\ee
and the result is
\bitem
\item
We look only at the sectors that are proportional to 
$(C \x^2 C) \f_{\a}$  
\item
We do not look beyond expressions with no derivatives.
\item
We take    either  $\y$ or  $C$ to be present, but only one of them.
\item
An arbitrary number of  $\A$ (with no derivatives) may be present.
\eitem

\refstepcounter{orange}
{\bf \theorange}.{\bf \;Now it is easy to verify that the limited sector is indeed in $E_2$}
For our limited purposes we can write 
\be
d_1 = C^{\a} \na_{\a} + \oC^{\dot \a} \ov\na_{\dot\a} 
\ee
The space $E_1$ is easily shown to be a function only of the fields $\A,\A_{\a \dot \b},\A_{(\a \b), (\dot \a\dot \b)}$ and $ \y_{\a},\y_{(\a \b), \dot \b},\y_{(\a \b \g), (\dot \a\dot \b)},\cdots$ and their complex conjugates.

This means that, for present purposes, we can truncate $\na_{\a}$ to
\be 
\na_{\a}= \y^a_{\a} A^{a \dag} +\y^a_{(\a \b) \dot \b} A_{\b \dot \b}^{a \dag} + \cdots +  
\A_{\a \dot \b}^{a } \oy^{a\dag}_{\dot \b }+  
\A_{(\a \b) (\dot\a\dot \b)}^{a } \oy^{a\dag}_{\b   (\dot \a \dot \b) }+ \cdots
\la{naform}
\ee 
and neglect the terms $\cdots$.

We can ignore all  the terms in $d_1$ because we only consider $(C \x^2 C)  ( \A)^n (\y\; {\rm or}\; C)$  or its complex conjugate  $(\oC \x^2 \oC) ( A)^n  (\oy\; {\rm or}\; \oC)$ in $E_1$.  Also note that the terms 
$(C \x^2 C) $ are not connected with any other terms of the structure, from \ci{dixmin}, used in (\ref{structure}), by increasing or decreasing the number of $C$ ghosts, except for the terms $(C \x^2 C) C_{\a}$.  But the terms $(C \x^2 C) $ have no $A^a,\oy_{a \dot \b}$ in them, by assumption here.  They have only $\A_a,\y^a_{ \b}$, so $\na_{\a}$ in  (\ref{naform})
 cannot affect them (since it goes with a $C^{\a} $ but it has the wrong combination of fields to affect them), which means that
\be
d_1  (C \x^2 C) \f_{ \a}  T_a[\A] \y^{a \a} =0\;{\rm and}\; 
d_1^{\dag}  (C \x^2 C) \f_{ \a}  T_a[\A] \y^{a \a} =0
\Ra 
(C \x^2 C) \f_{ \a}  T_a[\A] \y^{a \a} \in E_2
\ee
\be
d_1  (C \x^2 C) 
T [\A]
 C^{ \a}=0 \;{\rm and}\;  d_1^{\dag}  (C \x^2 C) 
T [\A]
 C^{ \a}=0
\Ra 
(C \x^2 C) 
T [\A]
 C^{ \a} \in E_2
\ee
So now we need to look at $d_2$ in this sector.

\refstepcounter{orange}
{\bf \theorange}.{\bf \;Operator $d_2$:}
\la{d2}
This comes from the following construction described in  \ci{dixspecseq}
\be
d_2 = \P_2 \d_1 \fr{\d_0^{\dag}}{\D_0}\d_1 \P_2 
\ee
where we need to take parts from (\ref{thisone1}), (\ref{deltazero})
and (\ref{thisone2}) for the three terms, respectively, as follows:
\be
d_2 = \P_2   \lt \{  \og^{ijk} \A_j \A_k {\ov \Lam}^{i\dag}   \rt \}
\lt \{  {\ov \Lam}^i F^{i \dag} \rt \} \lt \{  F^i C_{\a} \y_{\a}^{i \dag} \rt \} \P_2= 
\lt ( \og^{ijk} \A_j \A_k
    C_{\a} \rt )  \y^{i \dag}_{\a}
\ee

The part
$
\d_2 =  \x^{\mu} \partial_{\mu}  
$  has no effect on our limited study, which has no derivatives in it.
There are no further $d_r$ in this sequence. To see this requires some inspection along the lines in \ci{dixspecseq}, which we will not try to describe here.  So
\be
E_3 = \ker d_2 \cap  \ker d_2^{\dag} = E_{\infty} \approx \cH
\ee
and this is how we get the constraint equations, which come from:
\be
d_2 E_2 =  d_2^{\dag} E_2=0
\ee

\refstepcounter{orange}
{\bf \theorange}.\; 
\la{tricky}
 {\bf  Basic Remarks about the  isomorphism  $E_{\infty} \approx \cH$:}     
Once the elementary cases are understood, the rest follows easily, by simply multiplying each kind of expression by a product of $\A_i \cdots \A_n$ or  $A^i \cdots A_n$ depending on whether we start with $(C \x^2 C)$ or   $(\oC \x^2 \oC)$, respectively. It is easy to see how the  isomorphism $E_{\infty} \ra \cH$  works  for the simplest ghost charge zero free massless case:
\be
  (C \x^2 C)  
\A
  \ra 
\int d^4 x  {\ov F}    
\ee 
and   the simplest ghost charge one free massless case is not very different, when there is one free spinor index, as  discussed above in paragaph \ref{notaboundary1},  
 \be
  (C \x^2 C)  
\A
 C^{ \a}  \ra 
\int d^4 x  {\ov F} C_{\a}  
\ee
But it is surprising to see  the isomorphism which brings in the antifields and the anomalies  for the  simple free massless case, when there is one free spinor index,  as discussed above in paragraph \ref{simplestsourceofanom}.
\be
  (C \x^2 C)  
\y_{\a}
  \ra 
\int d^4 x  {\ov \G} C_{\a}  
\ee

 \refstepcounter{orange}
{\bf \theorange}.\; 
\la{trickysum}
 {\bf  Summary and Constraints:}    
Here is the more general isomorphism $E_{\infty} \approx \cH$ for this kind of exotic invariant and its anomaly.  The full superspace form of the exotic pair which can yield the anomaly  equation
\be
\d \cG[\cE_n] = ({\rm some\; coefficents}) \;\W_{n+2}
\ee
 is the following:
\be
\cE_n  =
T_{b}^{(a_1\cdots a_n)} \int d^4 x 
\lt \{ d^2 \oq 
\lt ( {\widehat \f}^{ \a} {\widehat \A}_{a_1} \cdots {\widehat \A}_{a_n} \; {\widehat {\ov \Lam}^b}
\rt ) C_{\a}  
+
  d^4 \q \lt ( {\widehat \f}^{ \a}  {\widehat \A}_{a_1} \cdots {\widehat \A}_{a_n} \;  {\widehat A}^b  \q_{\a} \rt )  \rt \}  
\la{omegaagainsupwithindices}
\ee 
\be
\W_{n+2}  =
T^{(a_1\cdots a_{n+2})} \int d^4 x  d^2 \oq 
\lt ( {\widehat \f}^{ \a}   C_{\a}  
   {\widehat \A}_{a_1}  \cdots  {\widehat \A}_{a_{n+2}}  \rt ) 
\la{anom33}\ee 
 We just need  to ensure that the  constraint equations are satisfied.  The mapping   $E_{\infty} \lra \cH$ is, for ghost charge zero $E^0_{\infty}\ra \cH^0$
\be
 (C \x^2 C) \f_{ \a}  T_{n,a}[\A] \y^{a \a} =\cE'_n \in E^0_{\infty}\ra \cE_n\in \cH^0
 \ee
and for ghost charge one $E^1_{\infty}\ra \cH^1$
 \be
(C \x^2 C) \f_{ \a}  T_{(n+2)}[\A] C^{\a} =\W'_{n+2}  \in E^1_{\infty}\ra \W_{n+2}\in \cH^1
 \ee
and the constraints  for these two kinds of expressions are as follows, for this massless case.  
\be
\og^{abc} T_{n,a}[\A] \A_b\A_c  =0
\la{anomeconstraintforcalE}
\ee
The n here means that there are n $\A$ fields in the expression $T_{n,a}[\A]$ and there are (n+2) $\A$  fields in $T_{(n+2)}[\A]$ fields so that one ends up with n $\A$ fields in the constraint:
\be
  g_{abc} \A_b^{\dag} \A_c^{\dag}   T_{n+2}[\A]  =0
\la{anomeconstraintforOmega}
\ee
The minimum n to generate triangle diagrams is n=1, as discussed in paragraphs \ref{simplestexoticpair} and \ref{ssmexample}.

\refstepcounter{orange}
{\bf \theorange}.\;
{\bf  Conclusion:}
\la{conclusion}
  These results generate some complicated problems.   The chief problem is choosing an exotic pair $(\cE'_n,\W'_{n+2})$  in a model with some Yukawa potential $\cP$ so that the two are likely to be connected by a set of triangle diagrams, generating a SUSY  anomaly $\d \cG[\cE_n] = ({\rm some\; coefficents}) \;\W_{n+2}
$.  Both need to be solutions of the anomaly constraint equations (\ref{anomeconstraintforcalE}) and (\ref{anomeconstraintforOmega}), or their massive analogues.    The exotic invariant $ \cE_n \in \cH^0$  will be  antifield dependent and superspace non-invariant;  and the corresponding potential anomaly $\W_{n+2}\in \cH^1$  will be proportional to the ghost $C_{\a}$. The most appealing case for the sake of simplicity is n=1.  We do know that for the massless case it is possible to generate solutions for both, for n=1, in the SSM, but they look unlikely to generate a SUSY  anomaly, unless there are masses present. That seems to point to  spontaneous gauge symmetry breaking. It is quite easy to implement that, and the result is that the exotic invariants disappear when spontaneous breaking occurs in the SSM, because they are no longer in the kernel of $d_2$, when that theory is massive.  So that indicates that  one  should be looking at higher $(A)^n$  values for the SSM, or at GUT theories \ci{ross} to find a SUSY anomaly. There may also be much simpler models with exotic pairs and SUSY anomalies. There is another serious issue.   Suppose that one finds an example of a SUSY  anomaly of this kind, after choosing some model and some superpotential $\cP$.  What does that mean?  Is it related to some kind of SUSY mass splitting that we could actually observe?  How does that work? Do these anomalies affect supergravity somehow \ci{west}?

\tiny
\articlenumber\\


\begin{thebibliography}{99}
\bibitem{susyconf}  There have been large conferences every year for a long time entitled `SUSY'.  Every year there are many talks on the SSM.
\end{thebibliography}
\end{document}